\documentclass[aps,prd,reprint,preprintnumbers,superscriptaddress,showpacs,twocolumn]{revtex4-1}
\usepackage{latexsym,graphicx,amssymb,amsmath,mathrsfs}
\usepackage{setspace,bm}
\usepackage[breaklinks, colorlinks=true, pdfstartview=FitV, linkcolor=red, citecolor=blue, urlcolor=blue]{hyperref}
\usepackage{amsmath}
\usepackage{ascmac}
\usepackage[usenames]{color}
\usepackage{latexsym}
\usepackage{epstopdf}
\usepackage{amssymb}

\newcommand\nc{N_\mathrm{c}}

\usepackage{ulem}
\usepackage{mathtools}

\renewcommand\sout{\bgroup \color{red} \ULdepth=-.5ex \ULset}

\begin{document}
%\preprint{****-**-**}

\title{Multiplicity, probabilities, and canonical sectors for the cold QCD matter}

\author{Kouji Kashiwa}
\email[]{kashiwa@fit.ac.jp}
\affiliation{Fukuoka Institute of Technology, Wajiro, Fukuoka 811-0295,
Japan}
\author{Hiroaki Kouno}
\email[]{kounoh@cc.saga-u.ac.jp}
\affiliation{Department of Physics, Saga University, Saga 840-8502,
Japan}

\begin{abstract}
At sufficiently low temperature, without requiring any numerical data at finite real chemical potential, we can clarify the canonical partition function with fixed quark number via the imaginary chemical potential region with few ansatzs.
The canonical partition function relates to the multiplicity distribution which can be observed in collider experiments and thus we may access important information of the properties of the QCD matter based on the canonical method.
In this paper, we estimate the multiplicity entropy, the configuration entropy, and the pointwise information which can be calculable with the canonical partition function to understand the properties of the cold QCD matter at finite density.
With the large $N_\mathrm{c}$ limit where $N_\mathrm{c}$ is the number of colors, we can simply estimate the tendency of them, and then the relation to the quarkyonic phase is clarified.
In addition, we discuss the nontrivial ground state degeneracy from the viewpoint of the canonical sectors.
\end{abstract}
\maketitle

\section{Introduction}

Exploring the phase structure of quantum chromodynamics (QCD) at finite temperature, $T$, and real chemical potential, $\mu=(\mu_\mathrm{R},0)$, is an important and interesting subject in elementary, hadron and nuclear physics and astrophysics.
There are several expectation of the QCD phase diagram based on lattice QCD simulations and QCD effective model calculations.
If we access the lattice QCD data at whole $\mu_\mathrm{R}$ region, we can have exact QCD phase structure.
It is, however, impossible at present because of the sign problem appearing at finite $\mu_\mathrm{R}$; see Ref.\,\cite{deForcrand:2010ys}.
The sign problem is weakened in the QCD effective models by simplifying the gauge field dynamics because the correlations between the gauge field and $\mu_\mathrm{R}$ relate to the seriousness of the sign problem.
However, QCD effective models have several unclearness in the foundation and then quantitative behaviors are not reliable.
Therefore, our understanding of the QCD phase diagram is strongly limited at finite $\mu_\mathrm{R}$.

Several phases are proposed so far on the QCD phase diagram at finite $\mu_\mathrm{R}$ such as the color superconducting phase, the inhomogeneously chiral symmetry broken phase such as the dual chiral density wave and the real kink crystal, the quarkyonic phase and so on; see Refs.\,\cite{Fukushima:2010bq,Buballa:2014tba} as an example.
These phases are deeply related to the $SU(N_\mathrm{c})$ color symmetry, the chiral $SU(N_\mathrm{f})_\mathrm{L} \times SU(N_\mathrm{f})_\mathrm{R}$ symmetry and the $U(1)_\mathrm{V}$ symmetry which relates to the quark number density where $N_\mathrm{c}$ and $N_\mathrm{f}$ are the number of colors and flavors, respectively.
In the standard approach, we construct suitable order parameters for the spontaneous symmetry breaking and clarify the phase structure. 
In Ref.\,\cite{Kashiwa:2021czl}, the authors showed that the fugacity expansion with the canonical partition functions are useful to clarify the properties of the cold QCD matter with sufficiently small $T$; particularly, the quarkyonic phase is expected to be deeply relate to the behavior of the canonical sectors. 

The quarkyonic phase is well defined in the large $N_\mathrm{c}$ limit, but it is not clear in the finite $N_\mathrm{c}$ system~\cite{McLerran:2007qj}.
The quarkyonic phase appearing at low $T$ is described as follows: 
The thermodynamics is dominated by the quark degree of freedoms under the Fermi surface and then the quark numbers start to have nonzero values ($\sim N_\mathrm{c}^1$) above the critical $\mu_\mathrm{R}$.
However, the physical excitation modes above the Fermi surface are baryonic (confined) in the large $N_\mathrm{c}$ limit.
The pressure is $\sim N_\mathrm{c}^0$ below the critical $\mu_\mathrm{R}$ because the glueballs are dominant degree of freedoms, but it turns into $\sim N_\mathrm{c}^1$ above the critical $\mu_\mathrm{R}$ because of the quark degree of freedoms under the Fermi surface.
In the deconfined phase, gluons become the dominant degree of freedoms and thus the pressure behaves $\sim N_\mathrm{c}^2$.
In addition, in the quarkyonic phase, the spatial inhomogeneity is expected to be appeared~\cite{Kojo:2011cn}; there are other types of inhomogeneity at finite density such as the dual chiral density wave~\cite{Nakano:2004cd} and the real kink crystal~\cite{Buballa:2014tba}.
These properties are estimated via the large $N_\mathrm{c}$ counting and thus they are not clear in the case of the finite $N_\mathrm{c}$.
However, some QCD effective model calculations indicate that the quarkyonic phase still exists in realistic QCD~\cite{McLerran:2008ua,Duarte:2021tsx} and it may directly affect the neutron star properties such as the mass-radius relation~\cite{McLerran:2018hbz}.
In addition, the quarkyonic phase may relate to the new confinement-deconfinement picture of QCD such as the soft-surface delocalization~\cite{Fukushima:2020cmk}.

In the quarkyonic phase, behavior of the baryonic modes plays a crucial role and thus it may be natural to think that the canonical partition functions have the hint to understand the phase because it has the relation to the multiplicity of baryon numbers.
However, this point is not well investigated so far.
It is well known that the canonical partition functions can be constructed via the imaginary chemical potential region~\cite{Roberge:1986mm,Hasenfratz:1991ax}; see Refs.\,\cite{Alexandru:2005ix,deForcrand:2006ec,Bornyakov:2017crk,Bornyakov:2016wld,Wakayama:2018wkc} for some progress.
In addition, the canonical partition functions relate to the multiplicity distribution which can be picked up from collision experiments; see Refs.\,\cite{STAR:2010mib,Luo:2012kja} as the experimental data and the recent review~\cite{Fukushima:2020yzx}.
For example, such fact was employed to connect the lattice QCD data and the experimental data  at finite density via the Lee-Yang zero analysis; see Refs.\,\cite{Nakamura:2013ska}.
The canonical partition functions can be constructed by using the imaginary chemical potential ($\mu_\mathrm{R}$), Fourier transformation and the fugacity expansion, and thus we can avoid the sign problem.
However, the sign problem is translated to the seriousness of the uncontrollability of the numerical accuracy of the Fourier transformation.
Even if the canonical method has the problem, the investigation of the QCD phase structure at finite $\mu_\mathrm{R}$ from the viewpoint of the canonical ensemble is interesting.

The purpose of this study is that we wish to consider what quantity is suitable to clarify the cold QCD matter at finite $\mu_\mathrm{R}$, particularly the quarkyonic structure, based on the quantity which can be observed in experiments and lattice QCD simulations.
Particularly, how we can account for the quarkyonic picture to clarify the QCD phase diagram at finite $\mu_\mathrm{R}$ based on the canonical sectors.
Therefore, this study has an impact on our understanding of the cold QCD matter at finite density which will be necessary to understand the neutron star properties.
This paper is organized as follows.
In Sec.\,\ref{sec:QCD} and \ref{sec:canonical}, we explain the procedure to construct the canonical partition functions.
The multiplicity distributions in the canonical and the grand canonical ensembles are explained in Sec.\,\ref{sec:canonical2} and \ref{sec:grand}.
Discussions are presented in Sec.\,\ref{sec:discussion}.
Section \ref{sec:summary} is devoted to summary.

\section{QCD with imaginary chemical potential}
\label{sec:QCD}

At finite pure imaginary chemical potential, $\mu=(0,\mu_\mathrm{I})$, QCD has several interesting and important properties.
We briefly summarize them below; see Ref.\,\cite{Roberge:1986mm,Kashiwa:2019ihm} for details.
\begin{description}
    \item[Roberge-Weiss (RW) periodicity]
    At finite imaginary $\mu$, several thermodynamic quantities and order parameters in QCD have the special $2\pi/N_\mathrm{c}$ periodicity along the $\theta$-axis where $\theta \equiv \mu_\mathrm{I}/T$.
    This special periodicity is the so-called the RW periodicity.
    Some details are shown in Appendix \ref{sec:app1}.
    \item[RW transition]
    The origin of the RW periodicity is different at low and high temperatures because of the balance between gluon and quark contributions in the grand canonical partition function.
    At $\theta=(2k-1)\pi/N_\mathrm{c}$ with $k \in \mathbb{Z}$, several quantities have singularities at high $T$.
    The $\mu$-odd and -even quantities have first-order and second-order singularities. 
    These singularities characterize the phase transition which is the so-called RW transition.
    \item[RW endpoint]
    Several quantities are smoothly oscillating along the $\theta$-axis at low $T$, but does not at high $T$.
    Because of the difference, there should be the endpoint of the first-order RW transition line.
    This endpoint is the so-called RW endpoint.
\end{description}

Since the imaginary $\mu$ can be translated into the temporal boundary condition of quarks, its effects must be vanished when we approach to the $T \to 0$ limit.
This means that the oscillating behavior of the grand canonical partition function as a function of $\theta$ can be assumed as
\begin{align}
    {\cal Z}_\mathrm{GC}(T,\theta)
    &= \sum_{n=0}^\infty a_n \cos (n\theta),
    \label{eq:dec}
\end{align}
and the coefficients of higher-order modes must be small, where $a_n$ must be real.
It should be noted that ${\cal Z}_\mathrm{GC}(T,\theta)$ is the $\theta$-even function and thus the $\cos$ function only appears.
At sufficiently low $T$, the constant mode $a_0$ becomes the dominant contribution, but higher-order oscillating modes will join the game when we increase $T$.
It should be noted that such the expansion is valid if there are no singularities along the $\theta$-direction.
In other words, such expansion can not work if there are singularities, which is known as the Gibbs phenomenon~\cite{gibbs1898fourier}.
Fortunately, we can think that there are no singularities along the $\theta$-axis at sufficiently low $T$ and thus we can employ this expression in this work. 

Since the expression (\ref{eq:dec}) is the Fourier decomposition, each oscillating mode, $a_n \cos (n \theta)$, is responsible to the corresponding canonical partition function, ${\cal Z}_\mathrm{C}(T,n)$.
Therefore, we have
\begin{align}
    {\cal Z}_\mathrm{C}(T,n) &\sim a_n,
\end{align}
when there are no RW transition in the $\theta$ region.
It should be noted that we only have $N_\mathrm{c} N$ contributions because of the RW periodicity where $N \in \mathbb{Z}$ which means the baryon number.
In other words, ${\cal Z}_\mathrm{C}(T,n)$ with $\mathrm{mod}(n,3) \neq 0$ becomes zero.
Since $a_{N_\mathrm{c}N}$ contribution represents the $N_\mathrm{c}N$ quark mode, the coefficient $a_{N_\mathrm{c}N}$ contains $\exp[-\beta N N_\mathrm{c} M]$ function; this form of the coefficient can be derived from the QCD effective model~\cite{Kashiwa:2021czl}.
Therefore, we work with the following expression;
\begin{align}
    {\cal Z}_\mathrm{C}(T,N_\mathrm{c}N) &\sim a_{N_\mathrm{c}N} e^{-\beta N_\mathrm{c}N M}.
\end{align}
and thus we have
\begin{align}
    {\cal Z}_\mathrm{GC}(T,\mu_\mathrm{R}) & \sim a_0 + \sum_{N=1}^\infty a_{N_\mathrm{c}N} e^{-\beta N_\mathrm{c}N(M-\mu_\mathrm{R})},   
\end{align}
where we neglect negative $N$ contributions because we only consider sufficiently large and positive $\mu_\mathrm{R}$ here.
It should be noted that the coefficient $a_{N_\mathrm{c}N}$ should have some suppression factor depending on $N$ because of the convergence which plays an important role as mentioned later.

\section{Canonical partition function}
\label{sec:canonical}

The canonical partition function with fixed quark number, ${\cal Z}_\mathrm{C}(T,n)$, can be constructed via the Fourier transformation of the grand canonical partition function at finite imaginary $\mu$, ${\cal Z}_\mathrm{GC}(T,\theta)$, by $\theta$ as
\begin{align}
    {\cal Z}_\mathrm{C}(T,n) &= \int \frac{d \theta}{2\pi} e^{i n \theta} {\cal Z}_\mathrm{GC}(T,\theta).
\end{align}
Then, the grand canonical partition function with real $\mu$ can be obtained with the fugacity, $\xi=\exp( \mu_\mathrm{R}/T)$, expansion as
\begin{align}
    {\cal Z}_\mathrm{GC}(T,\mu_\mathrm{R}) &= \sum_{n=-\infty}^\infty \xi^n {\cal Z}_\mathrm{C} (T,n). 
    \label{eq:canonical0}
\end{align}
Here we take the thermodynamic limit.
In the finite size system, the upper and lower bounds of the summation are restricted.
It plays an important role in the Lee-Yang zero analysis; for example, see Refs.\,\cite{Nakamura:2013ska,Nagata:2014fra} as an example.
It should be noted that each canonical partition function is independent with $\mu_\mathrm{R}$.
In the following, we use the canonical partition functions to clarify the QCD phase structure.

Since there is the RW periodicity, the canonical partition functions manifest the following relation;
\begin{align}
    &{\cal Z}_\mathrm{C} (T,n)
    \nonumber\\
    &= \frac{1 + z^n + z^{2n}}{2\pi}
       \int_{-\pi/3}^{\pi/3} d\theta \,
       e^{i n \theta} {\cal Z}_\mathrm{GC} (T,\theta)
       \nonumber\\
    &=
    \begin{dcases}
    \frac{3}{2\pi}  
    \int_{-\pi/3}^{\pi/3} d\theta \,
       e^{i n \theta} {\cal Z}_\mathrm{GC} (T,\theta)
       & (n = 3N)\\
     ~0 & (n \neq 3N)
    \end{dcases}
    ,
    \label{Eq:canonical}
\end{align}
for $N_\mathrm{c}=3$ where $z=e^{2\pi i /3}$.
Therefore, there are only the baryonic contributions which are not necessary to be truly confined.
Another way to construct the canonical sectors are using the restriction of the integral range as
\begin{align}
    {\cal Z}'_\mathrm{C} (T,n) &= \frac{1}{2\pi}\int_{-\pi/3}^{\pi/3} d\theta \,
       e^{i n \theta} {\cal Z}_\mathrm{GC} (T,\theta),
\end{align}
where the integral range is limited to one period of $\theta$.
In Ref.\,\cite{Kashiwa:2019dqn}, it is shown that this restriction can exactly reproduce the correct results at least with $\mu_\mathrm{R}=0$.
In this case, $n \neq 3N$ contributions do not become zero.
In this study, we use the former way (\ref{Eq:canonical}).

\section{Entropy in canonical ensemble}
\label{sec:canonical2}

In this study, we concentrates on the following function;
\begin{align}
    d(x) &= - x \ln x,
\end{align}
where $0 \le x \le 1 $.
This function maps $x$ to $\mathbb{R}$ and manifests the Leibniz rule; $d(xy)=d(x)y+xd(y)$ for $0\le y \le 1$.
Then, $\sum_N d(p_N)$ becomes the Shannon information entropy when $p_N$ is the probability.
By preparing several different probability distributions, we can consider different entropies which are responsible to different faces of the system.

The simplest probability distribution in the canonical ensemble is
\begin{align}
    p_N &= \frac{P_N}{ \cal N},
\end{align}
with
\begin{align}
    P_{N} &= {\cal Z}_\mathrm{C}(T,N),~~~~
    {\cal N} = \sum_N P_N = {\cal Z}_\mathrm{GC}(T,0).
    \label{eq:p_cano}
\end{align}
It should be noted that $\sum_N p_N=1$ must be satisfied for any definitions of $p_N$. 
This is the simplest choice of the probability distribution in the canonical method.

Since we now have the probability distribution function, we can calculate the entropy determined as
\begin{align}
    S &= \sum_{N=-\infty}^\infty d(p_N).
\label{eq:multi}
\end{align}
In the present calculation of the entropy, we individually consider all baryon number contributions.
In the experiments, if we prepare the distribution function by dividing the particle number histogram to each bin for the observed particles, we can have the probability distribution which has the relation to the above probability distribution (\ref{eq:p_cano}) because the canonical partition functions are directly related to the particle number if we focus on baryons.
Then, we can calculate the entropy via the same procedure.
This type of entropy is similar to the entropy which is the so-called the {\it multiplicity entropy}~\cite{Ma:1999qp}; for example, some details of several entropies are summarized in Ref.\,\cite{Li:2019uaq}.

In contrary, we can prepare another entropy based on the multiplicity distribution by considering the following setting;
\begin{align}
    S_{\mathrm{conf.},N} &=
    d(p_N) + d(p_\mathrm{other}),
\label{eq:conf}
\end{align}
with
\begin{align}
    p_\mathrm{other} = 1 - p_{N}.
\label{eq:setting}
\end{align}
In this entropy, we only focus on the one particular quark number contribution.
We call it {\it configuration entropy}.
Originally, the configuration entropy which is proposed in Ref.\,\cite{Li:2019uaq}, inspired by Ref.\,\cite{csernai2017quantitative}, is considered by focusing on one special internal configuration of proton.
In this sense, we focus on one special quark number configuration.

Unfortunately, the above entropies are independent of $\mu$ because they are only based on the canonical partition function and thus it is difficult to clarify the QCD phase structure at finite $\mu$.
Actually, in the experimental data at finite density, the multiplicity distribution must contain the $\mu$ effects.
Therefore, we consider the probability distribution explained in the next section.

\section{Entropy in grand canonical ensemble}
\label{sec:grand}

To understand the dense QCD matter, the probability distribution,
\begin{align}
    \tilde{p}_{N} &= \frac{\tilde{P}_N}{{\cal Z}_\mathrm{GC}(T,\mu_\mathrm{R})},
\end{align}
is a convenient probability distribution because it has the $\mu$-dependence where
\begin{align}
    \tilde{P}_{N} &= {\cal Z}_\mathrm{C}(T,N)\, \xi^N = P_N \xi^N.
\end{align}
Actually, this is the multiplicity distribution in the grand canonical ensemble.
Therefore, we can rewrote Eqs.\,(\ref{eq:multi}) and (\ref{eq:conf}) by replacing $P_N$ by $\tilde{P}_N$ and then we have $\mu$-dependent entropies.

Unfortunately,  Eq.\,(\ref{eq:multi}) with $\tilde{P}_N$ automatically contains all quark number contributions and thus it is not useful to clarify particular quark number contribution.
As mentioned in the introductory section, each quark number contribution may be important to understand the structure of the cold QCD matter at finite density and thus some other entropies which dissect the canonical sectors are desired.
In contrast with the thermodynamic entropy, Eq.\,(\ref{eq:conf}) with $\tilde{P}_n$ are matched with our purpose;
we call it the {\it $\mu$-dependent configuration entropy} in this study.
Particularly, the entropy based on the $N_\mathrm{c}$ quark contribution,
\begin{align}
    \tilde{S}_\mathrm{conf.,1} &= d(\tilde{p}_1) + d(\tilde{p}_\mathrm{other}).
\end{align}
can be expected to relate to the quakyonic phase transition which is well defined in the large $N_\mathrm{c}$ limit.
It should be noted that we need ${\cal Z}_\mathrm{GC}(T,\mu_\mathrm{R})$ to construct the distribution of $\tilde{p}$, but it is usually difficult.
To complete the construction, one way is to introduce the cutoff of $|N|$ or assuming the exponential suppression of the tail of the multiplicity distribution because we may expect the realization of the Gaussian distribution.
The former way is naturally implemented in the lattice QCD simulations because the lattice sites are limited as finite.
These treatments do not induce the serious problem until $\mu_\mathrm{R}$ approaches to the order of $\Lambda N_\mathrm{c} \Lambda_\mathrm{QCD}$, where $\Lambda_\mathrm{QCD}$ means the energy scale of QCD and $\Lambda \in \mathbb{Z}$ is the cutoff value of $|N|$, because their contributions are strongly suppressed.
If $\mu_\mathrm{R}$ goes over the value, inaccurate higher-order modes will affect the entropy.
This point is also true for some other entropies discussed in this paper.

In addition, the competition between each canonical sector may play a crucial role to clarify the dense QCD matter from the viewpoint of quarkyonic picture, and thus we here consider the following entropy;
\begin{align}
    D_{l} &= \sum_{N=-\infty}^\infty \tilde{q}_N \ln \frac{\tilde{q}_N}{\tilde{p}_N} = -\ln \tilde{p}_l,
    \label{eq:infor}
\end{align}
where $\tilde{q}=\{\cdots,\tilde{q}_{-1},\tilde{q}_0,\tilde{q}_1,\cdots\}$ with $\tilde{q}_l=1$ and the others are set to zero to sketch the shape of a particular baryon number ($l$) contribution.
This is based on the {\it Kullback-Leibler divergence} which is sometimes called the {\it relative entropy}~\cite{rached2004kullback}; it represents how similar the probabilities ($\tilde{p}$ and $\tilde{q}$) are.
It should be noted that $D_l$ corresponds to the pointwise information in the present setting of $\tilde{q}$ which is the important building block of the Shannon information entropy.

\section{Discussion}
\label{sec:discussion}

In this section, we show discussions on the qualitative behavior of entropies and some insights on the nontrivial ground state degeneracy from the viewpoint of the canonical sector, which is expected to have close relation with the confinement and deconfinement states.

\subsection{Qualitative behavior of entropies}
To estimate the behavior of Eq.\,(\ref{eq:infor}), we start from the expression of ${\cal Z}_\mathrm{GC}(T,\theta)$ at sufficiently low $T$ because we can image the behavior of ${\cal Z}_\mathrm{GC}(T,\theta)$ there, model-independently.
At sufficiently low $T$, the oscillation of ${\cal Z}_\mathrm{GC}(T,\theta)$ along the $\theta$-direction is strongly suppressed because $\theta$ corresponds to the phase for the  temporal boundary condition of quarks and it cannot affect the system at zero $T$.
Therefore, the contributions can be expected as
\begin{align}
    &{\cal Z}_\mathrm{GC}(T,\mu_\mathrm{R}) 
    \nonumber\\
    &\sim a_0 + \frac{1}{N_\mathrm{c}!}\tilde{a}_{N_\mathrm{c}} e^{-\beta N_\mathrm{c}(M-\mu_\mathrm{R})} + \frac{1}{(2N_\mathrm{c})!} \tilde{a}_{2N_\mathrm{c}} e^{-2\beta N_\mathrm{c}(M-\mu_\mathrm{R})},
\end{align}
where $\tilde{a}_n$ are coefficients where the suppression factor is extracted from $a_n$, and here we already neglect negative baryon number contributions because we are interested in $\mu_\mathrm{R} \sim N_\mathrm{c}M$.
It should be noted that more strong suppression can be expected at least at nonzero $T$; for example, see Re~\cite{Nagata:2011yf}. 
In this study, we assume the $1/n!$ suppression factor, but the stronger suppression factor does not change the following discussion.
Therefore, we can assume the probability distribution as
\begin{align}
    \tilde{p}_n = \frac{1}{n!} \frac{ \tilde{a}_n \, e^{- n \beta N_\mathrm{c}(M-\mu_\mathrm{R})} }{\cal N'},
\end{align}
where ${\cal N'}=\sum_n \tilde{p}_n$.
It should be noted that we assume that $1/n!$ suppression factor appears as the coefficient; this factor is natural to ensure that the grand canonical partition function does not diverge.
Another choice is the factor $1/n^s$ where $s$ is an exponent which characterizes the distribution; this form is based on the Zipf’s law~\cite{zipf2016human} which is an empirical law known in mathematical statistics.
Moreover, we may assume the Skellam distribution~\cite{jg1946frequency} based on the assumption that the proton and anti-proton follow the independent Poisson distribution.
Even in those three different forms, we can expect the additional suppression factor and thus the following discussions are unchanged if we take any of them.

For sufficiently small $\mu_\mathrm{R}$, $p_0$ dominates the probability distribution and thus Eq.\,(\ref{eq:infor}) with $l=1$ must be $\sim 0$.
In addition, higher-order $\tilde{p}_l$ with $l > 1$ will dominate the probability distribution when $\mu_\mathrm{R}$ departs from $\mu_\mathrm{R} \sim N_\mathrm{c}M$.
Therefore, Eq.\,(\ref{eq:infor}) with $l=1$ must have the peak structure at $\mu_\mathrm{R} \sim N_\mathrm{c}M$ at least in the large $N_\mathrm{c}$ limit.
For higher-order components ($D_l$ with $l > 1$), the position of the peak is not trivial because we should be careful with the competition between the exponential factor and the suppression factor. 
In the finite $N_\mathrm{c}$, particularly the sufficiently small $N_\mathrm{c}$, the peak structure even in $D_1$ should be smeared and becomes broad, but it may still have the information of the quarkyonic phase.
In addition, the configuration entropy is also expected to have the peak structure because of its functional form; $\tilde{p}_\mathrm{other}$ is almost zero when $p_{N_\mathrm{c}}$ dominates the probability distribution in the large $N_\mathrm{c}$ limit.
The appearance of the peak structure at the quarkyonic transition point may be similar to the situation for the nuclear liquid-gas transition discussed in Ref.\,\cite{Ma:1999qp}.

\subsection{Roberge-Weiss transition}
Finally, we discuss the relation between the present observation based on the canonical sectors and the nontrivial ground state degeneracy via the existence of the RW transition.
In Ref.\,\cite{Sato:2007xc}, it is shown that the confined and deconfined states at zero temperature are clarified from the nontrivial ground state degeneracy which has the close relation with the topological order~\cite{Wen:1989iv}; there is the nontrivial ground state degeneracy for the deconfined state but not for the confined state.
The discussions are proceeded on the compactified space-time by following three operations, the winding of the quark along the compactified dimension, the exchanging of quarks, and the insertion of the $U(1)$ flux to the hole which induces the Aharonov-Bohm phase; all operations must be done adiabatically.
In this study, we consider a sufficiently small $T$, where the Polyakov loop can be set to zero, and thus we may assume that we can approximately use the discussion shown in Ref.\,\cite{Sato:2007xc}.
Actually, we can introduce the flux insertion operation to the hole of the imaginary time direction and then the Aharonov-Bohm phase appears as the dimensionless imaginary chemical potential, $\theta:=\mu_\mathrm{I}/T$; for example, see Ref.\,\cite{Kashiwa:2015tna}.

At sufficiently low $T$ with $\mu_\mathrm{R}=0$, $\theta=\pi/3,\pi,5\pi/3$ which are changed to each other by the ordinary $\mathbb{Z}_3$ transformation, are belonging to the same state where $\theta$ acts as the trivial Aharonov-Bohm phase because the shift symmetry constructed by the semi-direct product~\cite{Kashiwa:2012xm,Shimizu:2017asf,Nishimura:2019umw} is not broken there; it is not possible if there is the RW transition because the shift symmetry is spontaneously broken.
In this sense, if the RW periodicity appears but not the RW transition in the system, the nontrivial ground state degeneracy cannot be expected in the $\theta$-direction.
Actually, the RW transition is induced by the quark contributions and thus the existence of the RW transition can be considered as an indicator to detect how strong the quark contributions against the baryonic contributions are;
for example, the strong coupling limit of QCD which is always in the confined phase does not have the RW transition~\cite{Roberge:1986mm,Kashiwa:2019ihm}.
Therefore, we may clarify the confinement and the deconfinement nature via the existence of the RW transition~ \cite{Kashiwa:2015tna,Kashiwa:2016vrl,Kashiwa:2017yvy}.

From the canonical approach, we can easily understand the existence of the Roberge-Weiss periodicity at finite $\mu_\mathrm{R}$ because the canonical partition functions seems to be free from $\mu_\mathrm{R}$ and thus we may simply complexify the fugacity as
\begin{align}
    \xi^N \in \mathbb{R} \to e^{ N (\beta \mu_\mathrm{R} + i \theta)} \in \mathbb{C},
    \label{eq:com}
\end{align}
to complexify the chemical potential in the grand canonical partition function.
In the following discussions, we are working with sufficiently low $T$, correctly speaking we are working with the almost $T \to 0$ limit, and thus we can evaluate ${\cal Z}_\mathrm{GC}$ based on the perturbation.
Therefore, we may rewrite ${\cal Z}_\mathrm{GC}$ as
\begin{align}
    {\cal Z}_\mathrm{GC}(T,\tilde{\mu})
    &\sim {\cal Z}_\mathrm{GC}(T,\mu_\mathrm{R})
     + \varepsilon(T,\tilde{\mu})
    \nonumber\\
    &= \sum_{N=-\infty}^\infty e^{\beta \mu_\mathrm{R}N } {\cal Z}_\mathrm{C}(T,N) + \varepsilon(T,\tilde{\mu}),
%    &= \sum_{N=-\infty}^\infty e^{(\beta \mu_\mathrm{R}+i\theta) N } {\cal Z}_\mathrm{C}(T,N) + \varepsilon(T,\tilde{\mu})
\label{eq:pert}
\end{align}
which is valid when $\varepsilon \in \mathbb{C}$ can be treated as the perturbation against $\rho_\mathrm{I}(\theta)$ where
$\rho_\mathrm{I}$ is the imaginary part of the quark number density and $\varepsilon$ is the infinitesimal $2\pi/3$ periodic function where $\rho_\mathrm{I}=0$ at $\theta=0$:
the first and second terms are the origin of the constant and periodic contributions against $\theta$.
In other words, we decompose ${\cal Z}_\mathrm{GC}(T,\tilde{\mu})$ to the constant and periodic parts and then we consider the certain $T$ region where the periodic term becomes sufficiently small. 
The first term in Eq.\,(\ref{eq:pert}) is dominant term and it does not create the singularities at finite $\theta$ by definition.
Therefore, $\varepsilon$ can have the contributions of the quark fugacity and they can induce the RW transition at finite $T$ if the quark fugacity contributes to the system markedly.
Here, we can assume that $\varepsilon$ cannot create the singularities when we work in the energy regime where $\varepsilon$ can be treated as the perturbation.
Since the evaluation of $\varepsilon$ needs actual model, we show a simple estimation below.
If the value $\beta \mu_\mathrm{R} \rho_\mathrm{I}$ approaches to the value $\beta \mu_\mathrm{R} \rho_\mathrm{R}$, where $\rho_\mathrm{R}$ means the real part of the quark number density, our perturbative treatment will be violated; $\beta \mu_\mathrm{R} \rho_\mathrm{R}$ and $\beta \mu_\mathrm{R} \rho_\mathrm{I}$ characterize the energy scale of the system because $\rho_\mathrm{R,I}$ are related with $\mu_\mathrm{R,I}$ via the differential calculus. 
At moderate $\mu_\mathrm{R} \sim \Lambda_\mathrm{QCD}$, we can assume $\rho_\mathrm{R} \sim k_\mathrm{F}^3 \sim \mu_\mathrm{R}^3 \neq 0$ after appearing the Fermi surface where $k_\mathrm{F}$ means the Fermi momentum.
This means that the perturbative treatment is acceptable if $\rho_\mathrm{I} = \alpha \mu_\mathrm{R}^3 \ll \Lambda_\mathrm{QCD}^3$ is manifested based on the order counting where $\alpha \ll 1$ which depends on $T$.
Since $\rho_\mathrm{I}$ decreases with decreasing $T$ but $\rho_\mathrm{R}$ dose not, we can find the regime, $\rho_\mathrm{I} \ll \Lambda_\mathrm{QCD}^3$, where the perturbative treatment is applicable by tuning $T$.
Because of above reasons, we can expect that the RW transition cannot be appeared along the $\theta$-axis at moderate $\mu_\mathrm{R}$ with sufficiently small $T$.
The above reasons can be also understood from the simple effective model estimation; see Appendix~\ref{sec:app3}.

Based on the discussions presented above, we can assume that the grand canonical partition function at finite $\mu_\mathrm{R}$ at least near $\mu_\mathrm{R} \sim M \sim M_\mathrm{B}/N_\mathrm{c} \sim \Lambda_\mathrm{QCD}$ must have the RW periodicity but not the RW transition if $T$ is small enough where $M_\mathrm{B}$ is the lowest baryon mass.
This fact means that $\theta$ may not be the nontrivial Aharonov-Bohm phase and then the nontrivial ground state degeneracy may be absent at moderate $\mu_\mathrm{R}$ with sufficiently small $T$.
At high $T$, quark modes must join the game even if $\mu_\mathrm{R}$ is very small and then the present discussion must be modified.
Therefore, at sufficiently small $T$, we can expect that the confined state continues to moderate $\mu_\mathrm{R}$ if we clarify it by using the nontrivial ground state degeneracy via the existence of the RW transition.
However, this picture may not be correct because we cannot distinguish the confined $N_\mathrm{c}$-quarks contribution and the deconfined $N_\mathrm{c}$-quarks ($\mathrm{mod}(n,N_\mathrm{c})=0$) contribution: this may be a serious problem at sufficiently small $T$ because other explicit $\mathbb{Z_{N_\mathrm{c}}}$ symmetry breaking contributions are strongly suppressed and then we cannot clarify the confinement-deconfinement nature based on the contribution.
Actually, the explicit $\mathbb{Z_{N_\mathrm{c}}}$ symmetry breaking contributions can affect the system and induce the RW transition at high $T$ and then the existence of the RW transition indicates the deconfinement energy scale.
This means that the absence of the RW transition at sufficiently small $T$ is not a sufficient condition but a necessary condition for the existence of the nontrivial ground state degeneracy. 
It should be noted that the quark number density can take the large value unlike the nuclear matter and then the structural change of the canonical sectors can be happened at least in large $N_\mathrm{c}$ limit~\cite{Kashiwa:2021czl}. 
Therefore, the nontrivial ground state degeneracy based on the existence of the RW transition is not enough to clarify the quarkyonic picture at sufficiently small $T$ and thus we need careful analysis of the structural change of the canonical sectors.

Some readers may concern how the color superconducting affects the ground state degeneracy because the diquark condensation breaks the color symmetry spontaneously in the gauge fixed calculation, and it seems to affect the RW periodicity when we make the canonical partition functions.
The important point is that $\mu$ and $A_4$ appears as the combination $\mu - iA_4$ in the Nambu-Gor'kov spinor for quarks ($S$) at finite chemical potential, which is widely used to consider the color superconductivity, as
\begin{align}
\tilde{S}^{-1}
&=
\begin{pmatrix}
\gamma \cdot p - m + \gamma_0\bar{\mu} & \Delta \gamma_5 \tau_2 \lambda_2  \\
-\Delta^* \gamma_5 \tau_2 \lambda_2 & \gamma \cdot p - m  - \gamma_0 \bar{\mu}
\end{pmatrix}
,
\end{align}
where $\Delta$ is the diquark condensation which can include the explicitly symmetry-breaking external-field, $\bar{\mu} = \mu - i A_4$, $p$ denotes the four-dimensional momentum, $m$ means the bare quark mass matrix, $A_4$ is the temporal component of the gluon field, $\gamma_5$ represents the fifth Dirac $\gamma$ matrix, $\tau$ does the Pauli matrices for the isospin space and $\lambda$ does the Gell-Mann matrices for the color space; for example, see Refs.\,\cite{Roessner:2006xn,Huang:2004ik} for details.
The above expression can be obtained from the four-Fermi interaction model with the mean-field of the temporal component of the gluon field which comes from the one-gluon exchange interaction with the local ansatz and the Fierz transformation, but the combination $\mu - iA_4$ is model independent; for example, see Ref.\,\cite{Huang:2004ik}.
This combination is the RW periodic combination which can be understood from the extended $\mathbb{Z_{N_\mathrm{c}}}$ symmetry~\cite{Sakai:2008py}.
Therefore, we can still have the RW periodicity even if there is the diquark condensation.
In addition, some readers are interested in the question that the higher-order diquark corrections for the mean-field approximation can break the RW periodicity or not.
It is not so trivial, but this question can be answered as "no" at least in the random phase approximation level as shown in appendix~\ref{sec:app2}.
Therefore, the above discussions for the ground state degeneracy are expected to be valid even if we consider the diquark contributions; see Ref.\,\cite{Kashiwa:2021czl} for some details how we can treat the diquark condensation in the canonical method.

It should be noted that we can also consider the flux insertion to the spatial holes by compactifying the spatial dimensions; it was considered in the original discussion of the topological order.
In this case, above discussions are still manifested if we set a sufficiently large compactified length, $L \to \infty$, because we can prove that the RW periodicity appears along the Aharonov-Bohm phase induced by the flux insertions and there is no spatial RW transition according to the same discussions in the case with the imaginary chemical potential; see Appendix \ref{sec:app1} for the spatial Roberge-Weiss periodicity.
From the discussions on the grand state degeneracy, we can understood the importance of the canonical sectors for the understanding of the confinement-deconfinement nature of QCD.

\section{Summary}
\label{sec:summary}

In this study, we have investigated what quantity is suitable to investigate the cold QCD matter at finite density based on the information theoretical viewpoint.
We have constructed the probability distribution in the canonical and the grand-canonical ensemble and propose some entropies.
From the estimation with the large $N_\mathrm{c}$ limit, we have shown that the configuration entropy and the pointwise information can pick up the structural change of the canonical sectors.
We found that the peak structure can be found in the configuration entropy and the pointwise information when we focus on the $N_\mathrm{c}$-quark contribution in the multiplicity distribution.
This peak structure may be related to the quarkyonic phase transition in the large $N_\mathrm{c}$ limit and its remnant may be expected to be appeared in the realistic $N_\mathrm{c}$ system.
In addition, we have discussed the cold QCD matter from the viewpoint of the nontrivial ground state degeneracy with the canonical method.
At least near $\mu_\mathrm{R}\sim M_\mathrm{B}/N_\mathrm{c}$ where $\mu_\mathrm{R}$ is the real chemical potential and $M_\mathrm{B}$ means the lowest baryon mass, the quark number density can become $O(N_\mathrm{c}^1)$ in the large $N_\mathrm{c}$ limit above $\mu_\mathrm{R} \sim M_\mathrm{B}/N_\mathrm{c}$.
In this situation, $\mod(n,N_\mathrm{c})$-quark (baryonic) contributions only survive and then the nontrivial ground state degeneracy may not be observed if we clarify it via the existence of the RW transition.

It should be noted that the above discussions are valid for the $\mu_\mathrm{R}$-dependence with fixed sufficiently small $T$ because the canonical partition functions are independent with $\mu_\mathrm{R}$, but do not with $T$.
Therefore, the $T$-dependence is more difficult than the $\mu_\mathrm{R}$-dependence in the present approach; this point is usually opposite in some other approaches.
Actually, the  multiplicity can be measured from the collision experiments via event-by-event observations and thus this approach seems to be useful in the future; more strictly speaking, the net-proton multiplicity can be observed, but the net-neutron multiplicity can not.
Of course, the collision experiment under the cold condition is quite difficult because the thermal system easily appears in the collision region with increasing collision energy.
However, the intermediate $T$ system can be explored in the experiments and thus the present discussion may be applicable if peak structure becomes very broad.

The lattice QCD simulation is also difficult to construct the canonical partition function at sufficiently low $T$ at present.
However, the canonical partition function at low $T$ is successfully estimated in some QCD effective models~\cite{Wakayama:2020dzz,Wakayama:2019hgz} such as the Polyakov-loop extended Nambu--Jona-Lasinio model~\cite{Fukushima:2003fw}. 
Therefore, if we can not obtain the canonical partition functions of QCD itself in the near future, we can utilize the proposed entropies in this paper with the QCD effective model, and then we can partially clarify the quarkyonic properties of the cold dense QCD matter.
In addition, the partial deconfinement~\cite{Hanada:2019czd,Hanada:2020uvt,Hanada:2018zxn,Watanabe:2020ufk} has been proposed recently.
The partial deconfinement is characterized via the spontaneous (partial) gauge symmetry breaking based on the Gross-Witten-Wadia transition~\cite{Gross:1980he,Wadia:2012fr} and also the spontaneous breaking of global symmetries.
Its realization with sufficiently low $T$ at finite density will be interesting and important in the future because it can connect the completely confined and completely deconfined phases and then it may have direct relation to the quarkyonic picture.
In such the case, we may need the detailed distribution of the gauge field and thus it is not clear how the distribution relates to the canonical partition functions at present.

\begin{acknowledgments}
This work is supported in part by the Grants-in-Aid for Scientific Research from JSPS (No. 19H01898 and No. 20K03974).
\end{acknowledgments}

\appendix

\section{Spatial Roberge-Weiss periodicity}
\label{sec:app1}

It is well known that QCD has the Roberge-Weiss (RW) periodicity at finite $\theta$~\cite{Roberge:1986mm}.
We can also prove that QCD with the spatial chemical potential has the RW periodicity model independently.
The QCD grand-canonical partition function is
\begin{align}
{\cal Z}_\mathrm{QCD}
 &= \int {\cal D} A {\cal D} {\bar q} {\cal D} q
    ~e^{-S_\mathrm{QCD}},
\label{Eq:GCPF}
\end{align}
having the action
\begin{align}
S_\mathrm{QCD}
&= \int d \tau d^3 x
   \Bigl[ {\bar q} \Bigl( \gamma_\mu D_\mu + m
        \Bigr) q - i T \theta q^\dag q
\nonumber\\
& \hspace{1.7cm}
        + \frac{1}{4g^2} (F^a_{\mu\nu})^2
   \Bigr],
\end{align}
where $q$ is the quark field, $A$ does the gluon field, $m$ denotes the bare quark mass matrix, $g$ means the gauge coupling constant, $D_\mu = \partial_\mu + i A_\mu$ and $F_{\mu\nu}$ represents the field strength tensor.
Here, we neglect the ghost and gauge fixing terms because these do not change following discussions.
Usually, we impose the antiperiodic boundary condition for the quark field in the temporal direction and then other boundary conditions are set to the periodic one because of the Fermi statistics.

The RW periodicity can be proven as follows~\cite{Roberge:1986mm}.
First, we redefine the quark field as
\begin{align}
 q (\tau, {\bf x}) \to e^{i T \theta \tau} q (\tau, {\bf x}).
\end{align}
This redefinition does not change the measure of the integral.
Then, the $\theta$-term vanishes from the grand canonical partition function, but the temporal boundary condition for quark fields is modified as
\begin{align}
 q(1/T,{\bf x}) = - e^{i \theta} q(0,{\bf x}).
\end{align}
Next, we use the $\mathbb{Z}_{\nc}$ transformation defined by
\begin{align}
 q &\to U_k q,~~~~
 A_\mu \to U_k A_\mu U_k^{-1} + i (\partial_\mu U_k) U^{-1},
\end{align}
with
\begin{align}
 U_k &= \exp \Bigl( i \frac{2 \pi k}{\nc} T \tau \Bigr).
\end{align}
The grand canonical partition function keeps its functional form, but the temporal boundary condition is changed into
\begin{align}
 q(1/T,{\bf x})
&= - \exp
   \Bigl[ i \Bigl( \theta + \frac{2 \pi k}{\nc} \Bigr)\Bigr]
   q(0,{\bf x}).
\label{Eq:BC1}
\end{align}
Thus, we can obtain the relation for the gran canonical partition function as
\begin{align}
{\cal Z}_\mathrm{QCD} \Bigl(\theta + \frac{2 \pi k}{\nc} \Bigr)
&= {\cal Z}_\mathrm{QCD} (\theta).
\end{align}
This relation is nothing but the RW periodicity.

If we compactify the spatial $x$-dimension and replace the chemical potential term in Eq.~(\ref{Eq:GCPF}) as
\begin{align}
 T \theta q^\dag q \to \frac{\theta_x}{L} {\bar q} \gamma_1 q,
\end{align}
we can also obtain the RW periodicity for $\theta_x$ where $L$ is the length of the compactified dimension.
We can interpret $\theta_x$ as the Aharonov-Bohm phase via the flux insertion to the loop; for example, see Ref.~\cite{huang1994statistical} for details.
Of course, $\theta_x$ is not conjugate valuable of the quark number operator ($q^\dag q$) and thus it is not the chemical potential, but we call $\theta_x$ the {\it dimensionless spatial imaginary chemical potential} for convenience.
Then, we can obtain the RW periodicity for $\theta_x$ as
\begin{align}
{\cal Z}_\mathrm{QCD} \Bigl(\theta_x + \frac{2 \pi k}{\nc} \Bigr)
&= {\cal Z}_\mathrm{QCD} (\theta_x).
\end{align}
by using the same procedure of $\theta$.
The boundary condition of quark fields for the $x$-direction becomes
\begin{align}
 q(\tau,L,y,z)
&= \exp
   \Bigl[ i \Bigl( \theta_x + \frac{2 \pi k}{\nc} \Bigr)\Bigr]
   q(\tau,0,y,z).
\label{Eq:BC2}
\end{align}
These relations are valid also at finite $T$. 
Extensions of the RW periodicity which corresponds to the $y$ and $z$ directions are straightforward.
The total sign of Eq.~(\ref{Eq:BC1}) and Eq.~(\ref{Eq:BC2}) is opposite because of the difference of the antiperiodic and periodic boundary conditions.
It should be noted that $\theta$ and $\theta_x$ are nothing but the boundary angle; this fact can be understood from Eq.~(\ref{Eq:BC1}), Eq.~(\ref{Eq:BC2}) and also the form of the frequencies.
%as shown later, namely Eq.~(\ref{Eq:MK}) and (\ref{Eq:MK2}).
The RW periodicity is deeply related to the center symmetry in the Yang-Mills theory and now the center symmetry is $(\mathbb{Z}_{\nc})_\tau \times (\mathbb{Z}_{\nc})_x$ at finite $T$ and $1/L$~\cite{Farakos:2004mm}.
Therefore, it is natural that the RW periodicity appears in both the thermal and spatial compactified dimensional chemical potential, $\theta$ and $\theta_x$.

To discuss the RW periodicities, the expectation values of the Polyakov loop and spatial Polyakov-loop operators, $\Phi$ and $\Phi_x$, are suitable quantities;
\begin{align}
 & \langle \Phi \rangle
 = \frac{1}{\nc} \Bigl\langle \mathrm{tr}_{\mathrm c}
               \Bigl[ {\cal P}
               \exp \Bigl( i\int_0^\beta A_\tau
               \hspace{0.5mm} d\tau\Bigr) \Bigr] \Bigl\rangle,
 \nonumber\\
 & \langle \Phi_x \rangle
 = \frac{1}{\nc} \Bigl\langle \mathrm{tr}_{\mathrm c}
               \Bigl[ {\cal P}
               \exp \Bigl( i\int_0^L A_x
               \hspace{0.5mm} dx \Bigr) \Bigr] \Bigl\rangle,
\end{align}
where ${\cal P}$ is the path-ordering operator.
With the antiperiodic thermal boundary condition, $\langle \Phi \rangle$ exist in the trivial center domain which means that the phase of the Polyakov loop ($\phi$) is zero, but it moves to the non-trivial center domain ($\phi \neq 0$) in the case of the periodic boundary condition because of the RW periodicity.
It is also true for the spatial boundary condition and the spatial Polyakov-loop.
This fact can be seen in Ref.~\cite{Ishikawa:2015nox} where the lattice QCD simulation is done in the torus system with the periodic spatial boundary condition and then the spatial Polyakov-loop exists in the non-trivial center domain.

In addition to the RW periodicity, the RW transition, which means the first-order transition described by the gap of the quark number density or the phase of the Polyakov loop, was also predicted in Ref.~\cite{Roberge:1986mm} and its prediction can be applied to spatial compactified dimension.
This fact can be easily understood from the supercubic symmetry of the Euclidean space-time; the RW periodicity and the transition in the $S^{1}_\tau \times R^3$ system is consistent with the spatial RW periodicity and the transition in the $S^{1}_x \times R^3$ system if we impose the same boundary condition to the $S^{1}_\tau$ and $S^{1}_x$ dimensions.
Then, the first-order spatial RW transition can be described by the gap of the $\langle {\bar q} \gamma_1 q \rangle$ or the phase of the spatial Polyakov-loop.

\section{Effective model estimation of RW transition}
\label{sec:app3}

In this Appendix, we discuss the system at sufficiently low $T$ with complex $\mu$ by using the Polyakov-loop extended Nambu--Jona-Lasinio (PNJL) model.
In the effective model computation, the partition function itself is difficult to compute, but we can image it via the effective potently which is corresponding to the grand (thermodynamic) potential.

The the Lagrangian density of the two-flavor PNJL model is given by
\begin{align}
{\cal L}
&= {\bar q} (i \gamma^\mu D_\mu - m_0 ) q + G[({\bar q}q)^2 + ({\bar q} i \gamma_5 \vec{\tau} q)^2]
 - {\cal U},
\end{align}
where $m_0$ denotes the current quark mass, $D_\mu$ is the covariant derivative $D^\mu = \partial^\mu + i g \delta^{4}_\mu A^\mu$ where $g$ denotes the gauge coupling constant, $G$ is the coupling constant and ${\cal U}$ is the gluonic contribution.
At least in the present discussion, ${\cal U}$ is not important and thus we do not show the explicit functional form of it; for example, see Refs.\,\cite{Fukushima:2003fw,Ratti:2005jh,Roessner:2006xn}.

The effective potential with the mean-field approximation can be expressed as
\begin{align}
    {\cal V} 
    &= -2 N_\mathrm{f}
       \int \frac{dp \, p^2}{4\pi^2}
       \Bigl[ N_\mathrm{c}E - T \sum_{\eta=\mp 1}\ln \det (1+e^{-\beta (E + \eta \mu')}) \Bigr]
    \nonumber\\
    & \hspace{0.4cm} + G \sigma^2 + {\cal U},
    \nonumber\\
    &= -N_\mathrm{f}
       \int \frac{dp \, p^2}{2\pi^2}
       \Bigl[ N_\mathrm{c}E + T \ln (f^- f^+) \Bigr]
        + G \sigma^2 + {\cal U},
\label{Eq:ep1}
\end{align}
where $N_\mathrm{f}=2$, $\mu' = \mu + i g A_4$ and
\begin{align}
    f^- &= 1 + N_\mathrm{c} (\Phi + {\bar \Phi}e^{-\beta E^- }) \, e^{-\beta E^- } + e^{- N_\mathrm{c} \beta E^- },
\nonumber\\
    f^+ &= 1 + N_\mathrm{c}  ({\bar \Phi} + \Phi e^{-\beta E^+ }) \, e^{-\beta E^+ } + e^{- N_\mathrm{c} \beta E^+},
\label{Eq:ep_PNJL}
\end{align}
here $E^\mp=E\mp\mu$, $\sigma = \langle {\bar q} q \rangle$ and $E=\sqrt{p^2+M^2}$ with $M=m-2 G \sigma$.
The Polyakov loop ($\Phi$) and its conjugate (${\bar \Phi}$) are then defined as
\begin{align}
    \Phi &= \frac{1}{N_\mathrm{c}} \mathrm{tr_c} \, e^{i\beta \langle g A_4 \rangle},~~
    {\bar \Phi} = \frac{1}{N_\mathrm{c}} \mathrm{tr_c} \, e^{-i\beta \langle g A_4 \rangle},
\end{align}
where $\mathrm{tr_c}$ is the trace acts on the color space.
The effective potential (\ref{Eq:ep_PNJL}) is nothing but the leading-order contributions in the $1/N_\mathrm{c}$ expansion scheme and then the higher-order meson-loop contributions are neglected.
However, we here concentrate on the quark contributions and thus this treatment can be acceptable.

Since we consider the sufficiently low $T$ region in the present study, the Polyakov-loop dynamics decouple from the system.
This means that we can explore the region where the Polyakov loop is sufficiently weaker than $\exp(\mp  \beta \mu_\mathrm{R})$, which are coupled with $\Phi$ and ${\bar \Phi}$ in the effective potential, by tuning the value of $T$.
If $\theta$-dependence of mean-fields is quite weak and then we can set them as a constant against $\theta$, we can simply discuss behavior of the effective potential as a function of $\theta$ as follows; this means that contributions of the $\theta$-dependence of mean fields can be evaluated as the perturbation.

The effective potential (\ref{Eq:ep1}) can be finally simplified as
\begin{align}
    {\cal V} 
    &= - N_\mathrm{f}
       \int \frac{dp \, p^2}{2 \pi^2}
       \Bigl[ N_\mathrm{c} E
            + T \ln ({\tilde f}^- {\tilde f}^+)
        \Bigr]
    + a,
    \label{eq:approx_eq_0}
\end{align}
where $a$ denotes the mesonic and gluonic parts which are $\mathbb{Z}_{N_\mathrm{c}}$ symmetric contributions and
\begin{align}
    {\tilde f}^\mp &= 1 + e^{- N_\mathrm{c} \beta (E \mp \mu) } + \epsilon,
\end{align}
here $\epsilon$ is the infinitesimal value which breaks the $\mathbb{Z}_{N_\mathrm{c}}$ symmetry explicitly.

If $\langle \Phi \rangle$ appears far from the origin in the complex $\Phi$ plane, the $\mathbb{Z}_{N_\mathrm{c}}$ symmetry breaking term, $\epsilon$, which rotates in the plane with varying $\theta$ can indices the RW transition, but does not if $\langle \Phi \rangle$ stays close to the origin; to appear the RW transition, clearly separated three minima, which are related to each other via the $\mathbb{Z}_\mathrm{c}$ transformation, are necessary. 
In the present study, we can prepare such a setting by tuning $T$ and thus we cannot expect the existence of the RW transition along $\theta$ even at finite $\mu_\mathrm{R}$; quark fugacities cannot affect the thermal system.
It should be noted that such the setting may be difficult to be prepared if $\mu_\mathrm{R}$ become much larger than $\Lambda_\mathrm{QCD}$ because several higher-order contributions which are neglected here may start to join the game markedly.

\section{Roberge-Weiss periodicity for diquark}
\label{sec:app2}

The detailed calculation method for the diquark polarization function  is almost the same to the mesonic case~\cite{Hansen:2006ee}.
Below, we do not consider the diquark condensate, but the consequence for the RW periodicity is unchanged because the combination of $\mu$ and $iA_4$ is the same.

The scalar diquark current is defined as
\begin{align}
    J_\Delta &= i q^T C \gamma_5 \tau_2 \lambda_A q,
\end{align}
where $C$ is the charge conjugation matrix and $\lambda_A$ are the antisymmetric components of the Gell-Mann matrices.
With the Wick contracting, we have the polarization function as
\begin{align}
    &\Pi_{\Delta}^\mathrm{s} (x) 
    \nonumber\\
    &:=\langle 0 | T(J_{\Delta}(x) J_{\Delta}^\dag(0) | 0 \rangle
    \nonumber\\
    &= \langle 0 | T(C \gamma_5 \tau_2 \lambda_A iS(x) \gamma_5 \tau^2 \lambda^A C^{-1} iS^T(x) ) | 0 \rangle.
\end{align}
Then, the polarization function in the momentum space becomes
\begin{align}
    \Pi_{\Delta}^\mathrm{s} (q) &= \int d^4 x \, e^{iq\cdot x} \Pi_{\Delta}^\mathrm{s}(x), 
\end{align}
Below, we set ${\bf q}=0$ for $q = q_0 + {\bf q}$.
By using the polarization function,
we obtain the following equation with random phase approximation (RPA) as
\begin{align}
&1-2G_\mathrm{d} \Pi^\mathrm{s}_\mathrm{\Delta} (q_0) \Bigl|_{q_0 = m_\mathrm{\Delta}} =0,
\end{align}
where $m_\Delta$ means the scalar diquark mass.
By using it, we can construct the diquark propagator in the RPA level.
The scalar diquark polarization function can be evaluated as
\begin{align}
&\Pi^\mathrm{s}_{\Delta}(q)
\nonumber\\
&= -\mathrm{tr_f} {\mathrm tr}_{\gamma}{\mathrm tr}_\mathrm{c} \int\frac{d^4p}{(2\pi)^4}
   [i\gamma_5 \lambda_A S(p-q) 
    i\gamma_5 \lambda_{A} S(p)]
\nonumber\\
&= - 4 \mathrm{N_f} {\rm tr}_\mathrm{c} \int \frac{d^4p}{(2 \pi)^4}
   \frac{p'p-M^2}
         {(p'^2-M^2)(p^{2}-M^2)} \lambda_A \lambda_{A} ,
\label{Eq:PF2}
\end{align}
where $p'= p+q$, $M$ is the constituent quark mass and $\mathrm{tr}$ acts in the corresponding parameter space.
It should be noted that $\mu$ and $A_4$ can be introduced by using the following replacement;
\begin{align}
p_0 &\to i \omega_n = i(2n+1) \pi T - q_0 - \mu + i A_4, 
\nonumber\\
p'_0 &\to i \omega_{n} = i(2n+1) \pi T + \mu - i A_4, 
\end{align}
and
\begin{align}
\int \frac{d^4p}{(2 \pi)^4} 
\to iT\sum_n \int \frac{d^3p}{(2 \pi)^3},
\end{align}
where the gauge coupling constant is absorbed into $A_4$.
%To easily evaluate the polarization function, it is good to decompose Eq.\,(\ref{Eq:PF2}) as
%\begin{align}
%   \Pi^\mathrm{s}_{\Delta}(q) &= 2 \mathrm{N_f}[A_1 + A_2 - q_0^2 B],
%\end{align}
%where
%\begin{align}
%    A_1 &= -i\mathrm{tr_c} \int \frac{d^4p}{(2 \pi)^4}
%         \frac{1}{p^2-M^2},
%    \nonumber\\
%    A_2(q_0) &= -i\mathrm{tr_c} \int \frac{d^4p}{(2 \pi)^4}
%         \frac{1}{p'^2-M^2},
%\end{align}
%and
%\begin{align}
%    B(q_0) &= -i\mathrm{tr_c} \int \frac{d^4p}{(2 \pi)^4}
%         \frac{1}{(p'^2-M^2)(p^2-M-2)}.
%\end{align}
After summing up all Matsubara frequencies, we will see the Fermi distribution functions and then we can evaluate the equation numerically if we set the model to calculate the constituent quark mass and the distribution of $A_4$.
Because of the functional form of $p_0$ and $p_0'$, the scalar diquark polarization function takes the different functional form about $\mu$ comparing with the pionic one; see Ref.\,\cite{Ishii:1995bu} for the coincidence between them at $\mu=0$.
We can clearly see the combination of $\mu-iA_4$ in $p_0$ and $p'_0$ appearing in the quark propagator and thus the above polarization function and also the diquark mass must have the RW periodicity.

\bibliography{ref.bib}

\end{document}